\documentstyle[12pt,epsf,twoside]{article}
\begin{document}

\font\names=cmbx10 scaled\magstep1
\newcommand{\be}{\begin{eqnarray}}
\newcommand{\ee}{\end{eqnarray}}
\newcommand{\Tr}{\mbox{Tr}}


\def\sf#1#2{\mbox{\small{$\frac{#1}{#2}$}}}
\def\f#1#2{\displaystyle\frac{#1}{#2}}    
\def\eq#1{Eq.~(\ref{#1})}
\def\cit#1{\cite{#1}}

\newcommand{\g}{\mbox{$\Gamma$}}
\newcommand{\gk}{\mbox{$\Gamma_k$}}
\newcommand{\gkp}{\mbox{$\Gamma'_k$}}
\newcommand{\ggf}{\mbox{$\Gamma_{\rm gf}$}}
\newcommand{\gktwo}{\mbox{$\Gamma^{(2)}_k$}}
%
\renewcommand{\i}{\displaystyle\int\!}
\def\mom#1{\!\displaystyle\frac{{d}^{d}#1}{(2\pi)^d}~}
\newcommand{\idx}{\int\!dx x^{\frac{d}{2}-1}}
\newcommand{\idthree}{\int\!dx~x^{\frac{1}{2}}}
\newcommand{\idd}{\int\!{d}^{d}x~}
\newcommand{\idm}{\int\!\frac{{d}^{d}p}{(2\pi)^d}~}
\newcommand{\Sum}{\displaystyle\sum}
%
\renewcommand{\d}{\delta }
\newcommand{\trave}{~---~}
%
\newcommand{\pd}[2]{\frac{\partial #1}{\partial #2}}
%
\newcommand{\fd}[2]{\frac{\delta #1}{\delta #2}}
\def\n{\noindent}
\def\l{\lambda}
\def\eb{\bar e^2}
\def\lb{\bar\l}
\def\e{{e^2}}
\def\ep{$U_k$~}
\def\r{\rho}
\def\o{\over}
\def\half{{1\o2}}
\def\F{\varphi}
\def\k{\kappa}
\def\L{\Lambda}
\def\tr{\tilde\rho}
\def\w{\omega}
\def\ss{\scriptscriptstyle}
%
%
\begin{titlepage}
\begin{flushright}
{HD-THEP-96-01}
\end{flushright}
\vspace{2.5cm}
\begin{center}
{\large\bf The Ward Identity from the Background Field}\\
\vspace{.1cm}
{\large\bf Dependence of the Effective Action}\\
\vspace{1cm}
{F. Freire\footnote{E-mail: F.Freire@ThPhys.Uni-Heidelberg.DE}
and C. Wetterich\footnote{E-mail: 
C.Wetterich@ThPhys.Uni-Heidelberg.DE}}\\
\vspace{.2cm}
{Institut  f\"ur Theoretische Physik, 
Universit\"at Heidelberg,}\\[.1ex] 
{Philosophenweg 16, D-69120 Heidelberg, Germany}\\
\medskip
\vspace{1cm}

\large\abstract{The dependence of the 
effective action for gauge theories 
on the background field obeys an exact identity. We argue that for
Abelian theories the Ward identity follows from the more general
background field identity. This observation is particularly relevant 
for the anomalous Ward identity valid for gauge theories with an 
effective infrared cutoff as used for flow equations.}

\end{center}
\end{titlepage}

\newpage

The effective average action $\gk$ is a 
useful concept for the investigation
of infrared problems. In field theory it corresponds to the quantum 
effective action with an effective infrared cutoff $\sim k$ for the
fluctuations. In statistical physics it is the coarse grained free
energy with coarse graining length scale $\sim$$k^{-1}$. 
Exact non-perturbative flow equations describe the dependence of 
$\gk$ on $k$ \cit1. They are related to the Wilsonian approach to the 
renormalization group equations \cit2. Since non-Abelian gauge theories 
are plagued in perturbation theory by severe infrared problems the use 
of the effective average action seems particularly promising here. 
Writing down exact flow equations for gauge theories poses no 
additional difficulty. The problem of correct implementation
of gauge symmetry arises rather on the level of their solutions. In fact,
models with local gauge symmetry correspond to particular trajectories in 
the space of general solutions of the flow equations. 
It is crucial to find 
appropriate identities for $\gk$ which enforce a restriction to 
``gauge-invariant solutions'' and guarantee gauge invariance for $k \to 0$.

So far, two different lines of research have been followed in this respect.
The first one \cit3\cit4 works in the background field formalism where
$\gk[A,\bar A]$  depends on the classical gauge field $A_\mu$
(conjugate to the source) and the background gauge field $\bar A_\mu$ 
which enters through the gauge fixing and the infrared cutoff. 
By construction $\gk[A,\bar A]$ is invariant under simultaneous 
gauge transformations of $A_\mu$ and $\bar A_\mu$.
This combined gauge invariance is, however, not sufficient to guarantee
that the solution lies on a trajectory appropriate for a gauge-invariant 
theory. An additional exact identity for the background field dependence
$\delta\gk/\delta\bar A_\mu$ was derived \cit4 and 
it is believed that this identity guarantees full gauge invariance of 
the theory. The second approach \cit5\cit6 centers on the 
Ward-Takahashi or Slavnov-Taylor identities.
These identities receive anomalous contributions \cit6 from the presence 
of the infrared cutoff term $\sim R_k$ which vanish only for $k \to 0$.
Obviously, the same type of identities can also be derived within the 
background field formalism. One would suspect that the background field
identity and the Ward identity are not unrelated  since both reflect the
content of gauge symmetry. Their exact relation has, however, not been 
revealed in the past. 

In this letter we indicate how the Ward identity can be derived from the
background field dependence of the effective action for Abelian gauge
theories. The model we use is scalar electrodynamics (SQED) in
arbitrary dimension $d$.
The classical action consists of the usual SQED action 
plus a gauge-fixing term and a quadratic term implementing an 
infrared cutoff. For the complex scalar field $\chi(x)$ the
infrared cutoff reads \cit4
\be
\Delta^{(S)}_k S = \idd \chi^* (x) R_k({\cal D}[\bar A])\chi(x).
\label{cutoff}
\ee
\n Here $R_k$ is a function that cuts off the modes with momentum 
smaller than $k$ with 
$R_k(0)\sim k^2$. More precisely, the cutoff distinguishes between
eigenvalues of the covariant Laplacian, ${\cal D}[\bar A] 
= - (\partial_\mu + ig \bar A_\mu(x))(\partial^\mu + ig\bar A^\mu(x))$,
where $\bar A_\mu(x)$ is the background gauge field. The complete set of 
quadratic cutoff terms includes a term for the gauge sector which can be 
chosen to be gauge invariant in Abelian gauge theories. If this is
the case it does not affect the Ward identity or the background field
identity and does not need to be specified for our discussion.

We choose the background gauge fixing
\be
\ggf[A, \bar A] = \f1{2\alpha} \idd 
\left(\partial_\mu[A^\mu(x)-\bar A^\mu(x)]\right)^2,
\label{gf}
\ee
and consider the effective action $\gk[\F,A,\bar A]$ with $\F$ the 
classical scalar field related to $\chi$. For this gauge fixing it has 
been shown that the background field dependence of $\gk$ is governed 
by the identity \cit4
\be
\f{\d\gk}{\d\bar A_\mu} = 
\Tr\left\{\f{\d {\cal R}^{(\F)}_k}{\d\bar A_\mu}~
\left(\gktwo+{\cal R}_k\right)^{-1}\right\} 
+\f1\alpha~\partial^\mu\partial_\nu(A^\nu-\bar A^\nu), \label{Aflow}
\ee
\n where the arguments of the functionals have been omitted. In 
\eq{Aflow} $\Tr$ stands for integration over the configuration 
or momentum space, e.g. $\Tr = {\displaystyle\idm}$ whilst 
$\left(\gktwo+{\cal R}_k\right)^{-1}$ denotes the inverse
of the two-point connected Green function for the complex scalar 
field in the presence of the infrared cutoff. 
In momentum space 
where $\left(\gktwo+{\cal R}_k\right)_{\F(p')\F(p)} =
\f{\d}{\d\F(p')} \f{\d}{\d\F^*(p)}\left(\gk+\Delta_k S\right)$,
we have

\be
\left(\gktwo+{\cal R}_k\right)^{-1}_{\F(p')\F(p)}=
\left<\chi^*(p')\chi(p)\right>_c.
\label{propagator}
\ee
By subtracting the gauge-fixing term
\be
\gkp[\F, A, \bar A] = \gk[\F, A, \bar A] 
- \ggf[A, \bar A], \label{newgamma}
\ee
\eq{Aflow} becomes
\be
\f{\d\gkp}{\d\bar A_\mu} = 
\Tr\left\{\f{\d {\cal R}^{(\F)}_k}{\d\bar A_\mu}~
\left(\gktwo+{\cal R}_k\right)^{-1}\right\}. \label{newAflow}
\ee

The Ward identity for SQED in the absence of cutoff functions is
well-known. In the presence of the infrared cutoff term 
$\Delta^{(S)}_k S$ it receives an anomalous contribution \cit6. For 
our choice of the gauge-fixing term the anomalous Ward identity reads
\be
\f1g~q_\mu \f{\d\gkp}{\d A_\mu(q)} +
\idm\left\{\f{\d\gkp}{\d\F(p)}\F(p-q) -
\f{\d\gkp}{\d\F^*(p)}\F^*(p+q)\right\} = \hspace{1cm}\nonumber\\[1.5ex]
\idm\frac{{\rm d}^{d}p'}{(2\pi)^d}~
\Big\{{\cal R}^{(\F)}_k(p+q,p')-{\cal R}^{(\F)}_k(p,p'-q)\Big\}~
\left(\gktwo+{\cal R}_k\right)^{-1}_{\F(p')\F(p)}.
\hspace{-1cm}\label{ward}
\ee
\n Note that ${\cal R}^{(\F)}_k(p,p')$ is not diagonal in momentum space 
($\partial_\mu$ and $\bar A_\mu$(x) do not commute) and depends on 
$\bar A_\mu$, \eq{ftexp}. The right-hand side of \eq{ward} vanishes
for $k=0\,\,({\cal R}_k=0)$ and the identity reduces to the standard
homogeneous linear relation between one-particle irreducible 
Green functions. In this case the identity is known to 
encode the transversality of 
the effects from fluctuations in the propagator of the gauge field
and it can be seen as a constraint guaranteeing the absence of
non-physical (in this instance longitudinal) degrees of freedom.
In the presence of the infrared cutoff the Ward identity
is a generalisation of this constraint. Truncated solutions 
to the exact flow equation should satisfy them at least approximately 
such that the homogeneous Ward identity is recovered for $k=0$. 

We now want to show that the Ward identity, \eq{ward}, follows from the 
background field identity, \eq{Aflow}.
The invariance of the background field effective action under a
simultaneous gauge transformation of $\F, A$, and $\bar A$ results in the
identity
\be
\f1g~q_\mu \f{\d\gkp}{\d A_\mu(q)} +
\f1g~q_\mu \f{\d\gkp}{\d\bar A_\mu(q)} +
\idm\left\{\f{\d\gkp}{\d\F(p)}\F(p-q) -
\f{\d\gkp}{\d\F^*(p)}\F^*(p+q)\right\} = 0. 
\label{totalward}
\ee
\n From \eq{totalward} the proof that \eq{ward} is contained in 
\eq{newAflow} follows immediately if we show that the following
equality holds
\be
\f1g~q_\mu\idm\mom{p'}\f{\d {\cal R}^{(\F)}_k(p,p')}{\d\bar A_\mu(q)}~
\left(\gktwo+{\cal R}_k\right)^{-1}_{\F(p')\F(p)}=\hspace{3cm}
\nonumber\\[1.5ex]
-\idm\frac{{\rm d}^{d}p'}{(2\pi)^d}
\displaystyle{\Big\{{\cal R}^{(\F)}_k(p+q,p')
-{\cal R}^{(\F)}_k(p,p'-q)\Big\}~
\left(\gktwo+{\cal R}_k\right)^{-1}_{\F(p')\F(p)}}. \hspace{-1cm}
\label{proof}
\ee
\n (Our computation in momentum space avoids problems due to operator 
ordering in $\f{\d {\cal R}_k[\bar A]}{\d\bar A_\mu}$.) We
need to evaluate ${\cal R}^{(\F)}_k(p,p')$ which corresponds to
$R_k({\cal D}[\bar A])$ in a 
Fourier basis $(\,\F(p)=\idd \F(x)\exp(ipx)\,)$
\be
\idd \F^*(x) R_k({\cal D}[\bar A]) \F(x) =
\idm\frac{{d}^{d}p'}{(2\pi)^d}~{\cal R}^{(\F)}_k(p,p') 
\F^*(p')\F(p). \label{ft}
\ee
For this purpose we 
expand $R_k$ in a power series of its argument
\be
R_k({\cal D}[\bar A]) = \Sum_{n=0}^{\infty}\f1{n!}
R_k^{(n)}(0)~({\cal D}[\bar A])^n, \label{power}
\ee
\n where $R_k^{(n)}(0) = \displaystyle{\f{{\rm d}^n R_k(x)}{{\rm d}x^n}
\Bigg\vert_{\,x=0}}$. Inserting
\be
{\cal D}(p,p')=p^2\bar\d(p-p')-g\i\mom{Q}\bar A^\mu(Q)(p+p')_\mu
\bar\d(p+Q-p')
\nonumber\\[1.5ex]
+\,g^2\i\mom{Q}\mom{Q'}\bar A^\mu(Q)\bar A_\mu(Q')~\bar\d(p+Q+Q'-p'),
\hspace{-1cm}\label{covder}
\ee
and comparing powers of $\bar A_\mu$ 
one arrives after a straightforward but lengthy 
manipulation at
\be
{\cal R}^{(\F)}_k(p,p') = 
R_k(p^2)\bar\d(p-p') -g\i\mom{Q}\bar A^\mu(Q)(p+p')_\mu
\f{R_k(p^2)-R_k(p'^2)}{p^2-p'^2}\bar\d(p+Q-p') \hspace{-.5cm}
\nonumber\\[1.5ex]
+\,g^2\i\mom{Q}\mom{Q'}\bar A^\mu(Q)\bar A^\nu(Q')~\bar\d(p+Q+Q'-p')~
\Bigg\{\d_{\mu\nu}\f{R_k(p^2)-R_k(p'^2)}{p^2-p'^2} \hspace{0cm} 
\nonumber\\[1ex]
-\f{(2p+Q)_\mu(2p'-Q')_\nu}{(p+Q)^2-p^2}~
\Bigg[~\f{R_k(p^2)-\f{p^2}{p'^2}R_k(p'^2)}{p^2-p'^2} \hspace{2.2cm}
\nonumber\\[1ex]
-\f{R_k((p'-Q')^2)-\f{(p'-Q')^2}{p'^2}R_k(p'^2)}{(p'-Q')^2-p'^2}~\Bigg]
\Bigg\} + {\cal O}(g^3), \hspace{.5cm}
\label{ftexp}
\ee
\n Here $R_k(p^2)$ equals $R_k(-\partial^2)$ in a momentum space 
representation and $\bar\d(p)=(2\pi)^d\d(p)$. We observe that 
${\cal R}^{(\F)}_k$ becomes diagonal only for $\bar A_\mu(x)$=const.

A direct computation of 
$\f{\d {\cal R}^{(\F)}_k(p,p')}{\d\bar A_\mu(q)}$ in 
order $g^2$ can be carried out using \eq{ftexp} and 
one wants to establish
\be
\f1g~q_\mu\f{\d {\cal R}^{(\F)}_k(p,p')}{\d\bar A_\mu(q)}=
{\cal R}^{(\F)}_k(p,p'-q)-{\cal R}^{(\F)}_k(p+q,p').   \label{result}
\ee
\n It is easy to verify this in lowest order. We have in addition checked
the first nontrivial order in $\bar A_\mu$ which is ${\cal O}(g)$. 
This shows that \eq{proof} holds
at least in this order, and we conclude that the Ward identity, \eq{ward}, 
is contained in the identity for the background field dependence of the 
effective action $\gk$, \eq{Aflow}.

We have no reasons to doubt that \eq{result} is an exact identity valid in 
all orders in $g$. It holds trivially if only constant background fields
are considered whereas a proof for general ${\bar A}_\mu(x)$ needs more
refined methods. We also expect that the background field identity implies
the Ward identity for all Abelian gauge theories with arbitrary matter
fields. This clarifies at least for the Abelian case the relation between
the two mentioned approaches to the use of exact flow equations for gauge
theories. In particular, the standard covariant gauge fixing is a special 
case of \eq{gf} for ${\bar A}_\mu(x)=0$. The background field identity, 
\eq{newAflow}, can also be used at ${\bar A}_\mu(x)=0$. 
For general $\bar A_\mu(x)$ 
\eq{newAflow} is stronger than the Ward identity since the 
latter is equivalent to its divergence, or, in momentum space, to a 
contraction with $q^\mu$. For practical applications one may isolate 
the gauge-invariant kernel ${\bar\Gamma}_k [\F,A]=\gk[\F,A,{\bar A}=A]$ and 
expand the remaining background field dependent part,
${\hat\Gamma}_k^{\rm gauge}[\F,A,{\bar A}]=
\gkp[\F,A,{\bar A}]-{\bar\Gamma}_k[\F,A]$, in powers of $A(x)-{\bar A}(x)$
\be
{\hat\Gamma}_k^{\rm gauge}[\F,A,\bar A]=\int d^{d}x
\Big\{H^\mu_1[\F,\bar A]\,\left(A_\mu(x)-\bar A_\mu(x)\right)\hspace{1.4cm} 
\nonumber\\[1ex]
+\half\,H^{\mu\nu}_2[\F,\bar A]\,\left(A_\mu(x)-\bar A_\mu(x)\right)\,
\left(A_\nu(x)-\bar A_\nu(x)\right)+\dots\Big\}.\hspace{-1cm}
\label{massterm}
\ee
Here $H_i$ are gauge-invariant Lorentz-tensors 
which may depend on $\F(x)$ and 
$\bar A_\mu(x)$ but not on $A_\mu(x)$. For any given ${\bar\Gamma}_k[\F,A]$ 
all $H_i$ can be computed from appropriate functional derivatives of 
\eq{newAflow} with $\gkp$ replaced by ${\hat\Gamma}_k^{\rm gauge}$. For 
example, a photon mass term for $\F=0$ is contained in
$H^{\mu\nu}_2=m_A^2\,\d^{\mu\nu}+\dots$\,.
The correction from ${\hat\Gamma}_k^{\rm gauge}$ to the photon two-point 
function reads
\be
\Delta G^{\mu\nu}(q)\bar\d(q-q')=-\left[
\f{\d^2{\hat\Gamma}_k^{\rm gauge}}
{\d\bar A_\mu(q)\d\bar A_\nu(-q')}
+\f{\d^2{\hat\Gamma}_k^{\rm gauge}}
{\d A_\mu(q)\d\bar A_\nu(-q')}
+\f{\d^2{\hat\Gamma}_k^{\rm gauge}}
{\d\bar A_\mu(q)\d A_\nu(-q')}\right]
_{\displaystyle 0},\,
\ee
where $[\dots]^{}_0$ means evaluated at $A=\bar A=\F=0$.
The photon mass term is extracted for $q=0$ ($\Delta G^{\mu\nu}(0)=
m^2_A \d^{\mu\nu}$) and one finds from \eq{newAflow} that it vanishes 
for $k\to0$ as $m^2_A\sim g^2k^2$. We conclude that 
${\hat\Gamma}_k^{\rm gauge}$ is completely fixed in terms of 
${\bar\Gamma}_k$ by the background field identity \eq{newAflow}\,\trave\, 
at least as long as the part ${\hat\Gamma}_k^{{\rm gauge}\,(2)}$ in 
$\g^{(2)}_k$ on the right-hand side can be treated iteratively. 

A background field identity has also been derived for non-Abelian gauge
theories \cit3 and up to now has been little exploited. It would be very
interesting to understand its relation with the anomalous
Slavnov-Taylor identities \cit6.
In view of the findings of this letter for the Abelian case one may suspect 
that the non-Abelian background field identity contains much relevant 
information beyond the Slavnov-Taylor identity.

\vfil
\newpage


\begin{thebibliography}{99}

\bibitem{1}
C.\,Wetterich, Z.\,\,Phys. {\bf C57} (1993) 451; {\bf C60} (1993) 461;\\
Phys. Lett. {\bf B301} (1993) 90;\\
M.\,Bonini, M.\,D'Attanasio and G.\,Marchesini,
Nucl.\,\,Phys. {\bf B409} (1993) 441.

\bibitem{2}
F.\,Wegner and A.\,Houghton, Phys.\,\,Rev. {\bf A8} (1973) 401;\\
K.\,G.\,Wilson and I.\,G.\,Kogut, Phys.\,\,Rep. {\bf 12} (1974) 75;\\
S.\,Weinberg, Critical Phenomena for Field Theorists, 
Erice Subnucl. Phys. (1976) 1;\\
J.\,Polchinski, Nucl.\,\,Phys. {\bf B231} (1984) 269.

\bibitem{3}
M.\,Reuter and C.\,Wetterich,  Nucl.\,\,Phys. {\bf B391} (1993) 147;
\newline{\bf B408} (1993) 91; {\bf B417} (1994) 181; preprint HD-THEP-94-39.

\bibitem{4}
M.\,Reuter and C.\,Wetterich,  Nucl.\,\,Phys. {\bf B427} (1994) 291.

\bibitem{5}
M.\,Bonini, M.\,D'Attanasio and G.\,Marchesini, 
Nucl.\,\,Phys. {\bf B418} (1994) 81; \newline{\bf B421} (1994) 429;
{\bf B437} (1995) 163; Phys.\,\,Lett. {\bf B346} (1995) 87.

\bibitem{6}
U.\,Ellwanger, Phys.\,\,Lett. {\bf B335} (1994) 364;\\
U.\,Ellwanger, M.\,Hirsch and A.\,Weber, preprint LPTHE Orsay 95-39,
to appear in Z.\,\,Phys. {\bf C}

\end{thebibliography}
\end{document}